\documentclass[
preprint,
aps,prd,
superscriptaddress,
showpacs,
tightenlines,
amsmath,
amssymb
]{revtex4}
\usepackage{bm}
\usepackage[dvips]{color,graphicx}
\newcommand{\ud}{\mathrm{d}}
\newcommand{\ksqmax}{k^2_{\mathrm{max}}}
\newcommand{\eq}[1]{Eq.(\ref{#1})}

\newcommand{\ceps}{\varepsilon}

\begin{document}

\title{Isospin Dependence of Power Corrections in Deep Inelastic Scattering.}

\author{S.I.~Alekhin}
\email[]{alekhin@sirius.ihep.su}
\affiliation{Institute of High Energy Physics, Protvino, Russia}

\author{S.A.~Kulagin}
\email[]{kulagin@ms2.inr.ac.ru}
\affiliation{Institute for Nuclear Research, 117312 Moscow, Russia}

\author{S.~Liuti}
\email[]{sl4y@virginia.edu}
\affiliation{University of Virginia, Charlottesville, Virginia 22901, USA.}


\begin{abstract}

We present results of a perturbative QCD analysis of  
deep inelastic measurements of both the deuteron and proton
structure functions. We evaluate the theoretical uncertainty
associated to nuclear effects in the deuteron, and we extract simultaneously   
the isospin depedendence of: {\it i)} the higher twists terms; {\it ii)} 
the ratio of the longitudinal to transverse cross sections, 
$R= \sigma_L/\sigma_T$;
{\it iii)} the ratio of the neutron to proton structure functions, 
$F_2^n/F_2^p$.
The extraction of the latter, in particular, has been at the center 
of an intense debate.
Its accurate determination is crucial both theoretically and 
for the interpretation of the more precise neutrino experiments
including the newly planned high intensity 50 GeV proton synchrotron.
\end{abstract}

\maketitle

\section{Introduction\label{intro}}


Deep Inelastic Scattering (DIS) experiments
provide the most accurate measurements of the strong coupling constant,
$\alpha_S,$ at intermediate scales. They are also the main source of information on
the Parton Distribution Functions (PDFs) in the proton and neutron.
The precision with which both of these quantities are known reflects directly into
the precision of calculations of the cross sections for 
all other hard scattering processes.
An accurate determination of these, in turn, plays a key role both in
the extraction of possible contributions
of new physics at new collider energies, and
in the interpretation of the
forthcoming high precision experiments using neutrino beams 
\cite{NuInt01}.

Both $\alpha_s$ and PDFs are not directly observable and
they need to be extracted from the DIS data according
to some procedure.
A number of uncertainties affect the analysis, related to both the
perturbative QCD (pQCD) series -- 
inclusion of higher orders, threshold resummation effects --
and to corrections that are non-perturbative in nature -- Target Mass Corrections (TMC),
dynamical power corrections, and nuclear effects in the case of the neutron
structure function.

It is therefore mandatory to be able to control the size of these
uncertainties by introducing a systematic, well tested, method of
extraction in which possible ambiguities can be properly gauged.
While analyses along these lines 
exist for the proton structure function 
(\cite{Ale1} in the DIS region and \cite{Liu1} in the ``few GeV'' or 
resonance region), an accurate and complete 
treatment of the neutron structure function
is still lacking.   
This paper is devoted to the application of a newly developed 
method to determine the isospin dependence of the  
nucleon Structure Functions (SFs). With the analysis presented here we hope to contribute  to
the interpretation of both recent data and new experiments,
by providing a quantitative measure of
the space in which pQCD based corrections and nuclear effects can be wiggled.

For our analysis we use the extensive DIS measurements that
exist for both proton and deuteron targets in a wide range of kinematics with the
exception for the very large $x$ region where fewer data sets exist, mostly at low
values of the final state invariant mass, $W^2$, in the region of
nucleon resonances (more experiments are however being planned that will cover the large $x$ DIS region
in forthcoming programs 
at Jefferson Lab \cite{jlab_proposals}, and at neutrino
facilities \cite{NuInt01}).

In QCD, different contributions to the DIS structure functions
can be written using the Operator
Product Expansion (OPE), by ordering them according to their twist, $\tau$
($ \tau =$ dimension $-$ spin) \cite{Wilson}.
%
The Leading Twist (LT)
contribution (with $\tau=2$ in DIS) is directly related to the
single particle properties of quarks and
gluons inside the nucleon - the PDFs.
The Higher Twist (HT) components ($\tau =4,6,...$ in unpolarized DIS) involve interactions between
quarks and gluons in the nucleon and they are suppressed by terms of order
$1/Q^2, 1/Q^4 ...$, respectively.
In phenomenological studies, the PDFs are extracted
from QCD global fits.
Accurate extractions use data with
sufficiently high $Q^2$ and invariant mass $W^2$, where
both target mass and HT corrections are expected to be very small.
QCD fits can now be performed to order $\alpha_S^3$ (Next-to-Next-to-Leading Order, NNLO
approximation). If the data encompass
a large range in $Q^2$, higher order corrections as well as HT
effects need to be taken into account simultaneously.

In a  recent series of papers \cite{Ale1} 
a proper choice of the statistical estimator allows
one to propagate all experimental error into the 
uncertainties in the PDFs. 
Because of the statistical efficiency of this new estimator,
the overall systematic error on the PDFs is sensibly
reduced with respect to previous analyses based on  
simplified estimators.
With a better determination of systematic experimental errors in hand, one can
address in detail the sources of theoretical errors.
Theoretical uncertainties are in principle an elusive concept
as by definition they refer to quantities that have yet to be calculated.
Uncertainties/ambiguities of this type and inherent to
the pQCD analyses are due to:
{\it i)} impact of the higher order QCD corrections;
{\it ii)} the HT terms; 
{\it iii)} target mass corrections (TMC);
{\it iv)} heavy quarks masses and threshold values; 
{\it v)} the form of the initial PDFs.
These questions were addressed in detail in the analysis of \cite{Ale1}, where
the DIS cross sections were fitted by including both the LT terms 
calculated from the PDFs evolved to the NNLO QCD, and the twist-4 terms, 
evaluated separately for the proton and neutron structure functions 
$F_2$ and $F_L$. 
The fact that the cross sections data
were fitted instead of the data on $F_2$, 
allowed for a better determination of $F_L$, which
in \cite{Ale1} was obtained iteratively.
In summary, the analysis of \cite{Ale1}
shows that theoretical uncertainties from the pQCD series are under control,
and that, due to the new estimator, all extracted quantities can be determined
with smaller errors than in previous analyses.

An additional uncertainty is however present in
analyses of the isospin dependence of both PDFs
and the HT terms, in that the neutron structure functions
have to be extracted from nuclear data. 
%
The main thrust of our analysis 
has been to make a thorough assessment of the impact of 
nuclear corrections on both the LT and HT terms.
Here, in particular, 
we focus on the isospin dependence of the HT terms.
Detailed results on both the ratio $F_2^n/F_2^p$ and on $R=\sigma_L/\sigma_T$
will be presented in a forthcoming paper \cite{AKL_future}. 
In our analysis we use the deuteron data where uncertainties are expected
to be in better control.
We address uncertainties arising from: {\it i)} Different models
of nuclear effects -- we highlight in particular the differences with 
using the extrapolation \cite{YanBod} of the nuclear density model of the EMC effect \cite{FS88}; 
{\it ii)} Different
deuteron wave functions derived from currently available NN potentials,
giving rise to different amounts of high momentum components;
{\it iii)} The interplay between nucleon off-shellness and TMC in nuclei.
%

In addition to its practical purpose, a quantitative determination of the isospin dependence
of the HTs contributions
is of theoretical interest in understanding the nature of power corrections.
On one side, infrared renormalons (IRR) have been suggested 
as a method for estimating the contribution of
power corrections to the cross sections for a number of hard processes
(see \cite{BenBra} and references therein).
Based on this hypothesis the calculations in \cite{DMW,SMMS} have
predicted the $x$ dependence
of the coefficients of the HT terms for both $F_L$ and the valence and singlet
contributions to $F_2$.
On the other hand, some models
exist that predict a sizeable isospin dependence of the HT terms. 
This has been suggested for instance in models that interpolate between partonic 
and non-partonic degrees of freedom at low $Q^2$ as in \cite{SzcUle} (a smaller 
effect seems to occur, however, in the predictions for the HT isospin 
dependence in \cite{CKMT}). A large effect of about a factor two for the ratio of the 
neutron to proton HT terms, was also predicted 
based on quark counting estimates in Ref.\cite{Bro1}.

A thorough analysis of the isospin dependence of HTs might therefore
help disentangling predictions of different models, by investigating, for
instance, whether the onset of  ``flavor-blind'' power corrections from IRR
occurs at a different scale than for other dynamical ones.


In what follows, we present our results for each quantity along with a
discussion of the nature of the nuclear effects and the extraction method.
In Section \ref{formal} we present the general formalism and definitions.
In Section \ref{nuclear-sect} we discuss the contribution of nuclear
effects. In Section \ref{fit} we outline the extraction method and 
we present our results. In Section \ref{pheno-sect} we discuss some phenomenological 
applications of our analysis. Finally, we draw
our conclusions in Section
\ref{conclusions}.


\section{DIS Formalism\label{formal}}

The inclusive DIS section of unpolarized charged leptons off
an unpolarized nucleon or nuclear target is fully determined
by the spin-averaged electromagnetic tensor of the target,
which can be parameterized in terms of two invariant structure
functions $F_1$ and $F_2$ (we ignore a small contribution due to neutral current)
\begin{eqnarray}
\nonumber
W_{\mu\nu}(p,q) &=& \frac1{8\pi}\sum_\mathrm{s}
\int \ud^4 z \exp(iq\cdot z)\,
\langle p,s| \left[J_\mu^\mathrm{em}(z), J_\nu^\mathrm{em}(0) \right]|p,s \rangle \\
\label{W:SF}
&=&{} -\widetilde g_{\mu\nu}F_1
+\frac{\widetilde p_\mu \widetilde p_\nu}{p\cdot q}F_2,
\end{eqnarray}
where $J_\mu^\mathrm{em}$ is electromagnetic current,
$p$ and $q$ are the target momentum and four-momentum transfer, respectively.
In order to simplify notations, we denote
\begin{eqnarray}
\label{g-tilde}
\widetilde g_{\mu\nu}&=&g_{\mu\nu}-\frac{q_\mu q_\nu}{q^2}, \\
\widetilde p_\mu &=&p_\mu-\frac{p\cdot q}{q^2}q_\mu .
\label{p-tilde}
\end{eqnarray}
The normalization of states adopted here is
$\langle p|p'\rangle = 2p_0(2\pi)^3\delta(\bm{p}-\bm{p}')$
for both the bosons and fermions.
With this normalization the structure functions $F_{1,2}$ are dimensionless.
They depend on two invariant variables, namely the Bjorken variable
$x=Q^2/2p\cdot q$ and the four-momentum transfer squared $Q^2=-q^2$.

The differential cross section in terms of the structure functions and
standard variables $x$ and $y=p\cdot q/p\cdot k$, where $k$ is the incoming
lepton four-momentum, reads
\begin{eqnarray}
\label{xsect}
\frac{\ud^2\sigma}{\ud x\ud y} =
\frac{4\pi\alpha^2}{Q^2 xy}
\left[
    \left(1-y-\frac{(Mxy)^2}{Q^2}\right)F_2 +
    \frac{1}{2}{y^2}\left(1-\frac{2m_l^2}{Q^2}\right) F_T
\right],
\end{eqnarray}
where $\alpha$ is the electromagnetic coupling constant, 
$M$ is the nucleon mass and $Q^2=2xy p\cdot k \,$, and $F_T=2xF_1$.
\footnote{We keep the lepton mass in \eq{xsect} for the sake of
completeness. Although this term is negligible in electron scattering, we
take it into account in our analysis of muon data}.

In QCD the contributions from different quark-gluon 
operators to the electromagnetic
tensor and to DIS structure functions
are ordered according to their twist, leading to the expansion in inverse
powers of $Q^2$:
\begin{eqnarray}
\label{t-expansion} F_{T,2}(x,Q^2) = F_{T,2}^{LT}(x,Q^2) +
\frac{H_{T,2}(x,Q^2)}{Q^2} + {\cal O}\left(1/Q^4 \right).
\end{eqnarray}
This expansion applies to both the proton and neutron structure functions, 
we have suppressed the indices for simplicity. 
The first term is the LT contribution and 
$H_{1,2}$ are 
the HT -- twist-4 -- contributions. Furthermore, if a finite mass for the
nucleon target is considered, new terms arise in  
Eq.(\ref{t-expansion}) that mix operators 
of different spin, leading to additional power terms of
kinematical origin -- the so-called target mass corrections.
In the approximation that $x^2M^2/Q^2$ is small, 
the TMC series can be absorbed in the leading twist term 
\cite{GeoPol76}.

The separation of LT, TMC and dynamical HT from the data, is not
straightforward as witnessed by the number of studies dedicated to it since
the initial formulation of the problem in the 1970's
\cite{GeoPol76,AbbBar}.
In what follows we define the LT, TMC, and
HT, contributions to the structure functions $F_T$ and $F_2$.
The formalism for nuclear DIS is discussed in Section \ref{nuclear-sect}. 

\subsection{Leading Twist
\label{LT-sect}}

The LT part of the structure functions is related to the PDFs, $p_i(x,Q)$ -- the index $i$ refers
to the different types of quarks and antiquarks, and to the gluon distribution -- 
via a convolution with perturbatively calculable
coefficient functions $C^i_{T,2}$:
\begin{eqnarray}
\label{LT_1}
F_{T,2}^{\mathrm{LT}}(x,Q^2) & = &  \sum_{i=q,\bar{q},g} \int_x^1\frac{\ud z}z
C_{T,2}^i(z,\alpha_S(Q^2)) p_i(x/z,Q^2).  
\end{eqnarray}
%

The $Q^2$ dependence of the PDFs is predicted by the well-known
DGLAP evolution equations \cite{DGLAP}:
%
\begin{eqnarray}
\label{DGLAP}
 t\frac{\displaystyle \partial p_i(x,t)}{\displaystyle \partial t}  =  \sum_{j=q,\bar{q},g} \int_x^1\frac{\ud z}z
P_{ij}(z,\alpha_S(t)) p_j(x/z,t), 
\end{eqnarray}
where $t=Q^2$, and $\alpha_S$ is the strong coupling constant, and $P_{ij}$
are the splitting functions.

The coefficient functions 
have been calculated to NNLO~\cite{C-NNLO}. 
The splitting functions are known to NLO, and only a limited set of
Mellin moments are evaluated to NNLO \cite{P-NNLO-moments}. Although, 
estimates of the full $x$-dependence of the splitting functions in NNLO
approximation are available \cite{P-NNLO}, in our analysis we use the 
$\overline{\mathrm{MS}}$ NLO QCD approximation with the
renormalization/factorization scales chosen equal to $Q$.
The NNLO variant of our fit is used to estimate the uncertainty
due to higher orders. 
Large $x$ resummation effects, arising from terms of the type 
$[\alpha_S(Q^2) \ln(1-z)]^{2k}$ at ${\cal O}(\alpha_s^k)$, in the coefficient functions
are present in principle. They have been shown to be comparable
in size to NNLO corrections in Ref.\cite{ScScSt}, and to generate a 
further negative correction to the HT coefficient of the proton SF.   


\subsection{Target Mass Corrections
\label{TMC-sect}}
We follow the method of Ref.\cite{GeoPol76} where it was shown that the 
TMC series can be summed up, leading
to the modification of the LT term. Therefore, Eq.(\ref{t-expansion}) remains valid
with the LT terms replaced by
\begin{subequations}
\label{TMC}
\begin{eqnarray}
F_T^{\mathrm{LT,TMC}}(x,Q^2) &=&
\frac{x^2}{\xi^2\gamma}F_T^{\mathrm{LT}}(\xi,Q^2) + 
2 \frac{x^3M^2}{Q^2\gamma^2}\int_\xi^1\frac{\ud\xi'}{{\xi'}^2}F_2^{\mathrm{LT}}(\xi',Q^2),\\
F_2^{\mathrm{LT,TMC}}(x,Q^2) &=&
    \frac{x^2}{\xi^2\gamma^3}F_2^{\mathrm{LT}}(\xi,Q^2) + 
    6\frac{x^3M^2}{Q^2\gamma^4}\int_\xi^1\frac{\ud\xi'}{{\xi'}^2}F_2^{\mathrm{LT}}(\xi',Q^2),
\end{eqnarray}
\end{subequations}
where $\gamma=(1+4x^2M^2/Q^2)^{1/2}$ and $\xi=2x/(1+\gamma)$ 
is the Nachtmann variable \cite{Nacht}.

It must be noted however, that the derivation of \cite{GeoPol76} was given
in the zeroth order in $\alpha_S$, assuming that the target quarks are
on-shell. Both, higher order $\alpha_S$ corrections and quark off-shell
effects modify Eqs.(\ref{TMC}). It was argued in \cite{FraGun80} that
off-shell effects lead to $M^2/Q^2$ terms which are not incorporated in
Eqs.(\ref{TMC}). In addition, 
target mass corrections should be applied also to the HT terms 
in the higher order terms in the expansion (\ref{t-expansion}). 
For this reason we do not consider $1/Q^4$ terms in the
TMC formula, which are small for the kinematical range considered. 
Finally, TMC corrections for 
an off-shell target, {\it i.e.} when $p^2\not = M^2$, 
should be treated as part 
of the nuclear effects and will be discussed therefore in Section III.

\subsection{Higher Twists \label{HT-sect}}
The extraction of higher twist terms from the data
is a longstanding problem, as recognized from    
the very first developments of a pQCD phenomenology
(\cite{DGP,AbbBar}).  
HTs have been hard to pin down for a number of reasons. 
First of all a connection with partonic interpretations cannot be established
on a one by one basis, differently from the LT components that 
are directly related to the PDFs. 
In fact the HT terms are formally written within OPE
as the product of coefficient functions and 
hadronic matrix elements of composite local operators. 
Not all of the matrix elements are independent, but a minimal basis can be
selected after relating them through the equations of motion
\cite{JafSol}.  
Nevertheless, even in a minimal basis, the number of independent 
reduced matrix elements is much larger than the number of observables
-- {\it e.g.} the moments of the structure functions. 
In unpolarized scattering one can single out formally the
four quarks and two quarks-two gluons types of operators, corresponding
in a partonic language to quark-quark and quark-gluon correlations, 
respectively. 
The determination of the relative scales of these contributions 
is at present model dependent. However, it was shown in Ref.\cite{Choi} that
a simultaneous analysis of both $F_2$ and $F_L$ better constrained 
the evaluation of the quark-gluon term, that was required to be large
and negative in order to fit both sets of data. 
 
A practical difficulty is also in that theoretical estimates 
-- from the simple estimate of the  
increase of the number of operators with respect to twist-2 \cite{DGP}, 
to more sophisticated IRR calculations \cite{DMW,SMMS} -- 
predict that HTs are most important at low $W^2 \approx Q^2(1-x)$. 
In this limit, with $W^2 \geq 4$ GeV$^2$, thus avoiding the resonance region, 
it can be easily
shown that the logarithmic dependencies characterizing pQCD evolution to 
a given order mimic the $1/Q^2$ dependence of the twist-4 terms.  
A correlation between the pQCD parameters and the HT coefficients 
arises that has lead the authors of \cite{Kataev:1999bp} to conclude that 
for the structure function $F_3$, the NNLO term and the HT corrections
are, within the precision of current data, undistinguishable.
A similar analysis was subsequently applied in Ref.\cite{YanBod} to
$F_2$. It lead the authors to the conclusion that HTs are highly 
reduced with respect to previous determinations, even inside the 
resonance region. 

To summarize, the results of a number of analyses of the HT 
contributions in DIS are still not conclusive. A joint search 
using a combination of predictions from hadronic models 
and an accurate experimental extraction seems to be the most promising avenue.
It is in this spirit that in this paper we address yet another aspect 
of the phenomenology of HTs, namely their isospin dependence. 
An isospin dependence of the HT terms resides entirely in their 
non-perturbative structure.   
In IRR models in fact the coefficients of the HT terms are predicted 
to have no  
target dependence, provided the radiative corrections to these terms
are factored out by assuming for instance, that they behave similarly to
their LT correspondent.
An assessment of the magnitude of the isospin dependence of the HT terms
provides therefore a handle on understanding their nature and more 
precisely the extent to which they can be described by models.
In particular, possible scenarios about the large $x$ 
structure of the proton can be investigated, envisaging coherent 
scattering from multiquark composites carrying increasing momentum
fraction at $x \rightarrow 1$ \cite{Bro1}.     

\section{Nuclear Effects} 
\label{nuclear-sect}
Since nuclear data are used in the QCD analysis, 
nuclear effects must be taken into account. 
In our analysis we use deuterium data in order to study
the isospin effect in PDFs and HT terms.
In this section we briefly discuss our method to correct for nuclear effects.
More details can be found elsewhere \cite{AKL_future}.
\subsection{Fermi motion and binding effects}
\begin{figure}
\begin{center}
%
\includegraphics{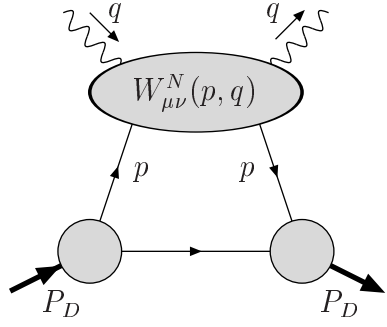}
\end{center}
\caption{Deuteron Compton scattering amplitude in incoherent scattering
approximation.}
\label{nucl_ia}
\end{figure}

For large $x>0.1$, away from nuclear shadowing region, nuclear DIS of
leptons off nuclear targets can be viewed as incoherent scattering off bound
nucleons \cite{West74,AKV85,Ku89,CL90,GL92,KPW94}.
The DIS cross section is given by the imaginary part of the virtual
photon Compton amplitude in the forward direction.
In incoherent scattering approximation, 
the nuclear Compton amplitude is taken in impulse
approximation by disregarding both initial state interactions and
interactions between the struck quark and the nuclear debris. The
corresponding Feynman diagram is shown in Fig.\ref{nucl_ia}. Calculating this
diagram and projecting the structure functions from the imaginary part of
the Compton amplitude we derive the relations between the nuclear and the
nucleon structure functions. For the
deuteron we have
\begin{subequations}
\label{D-SF}
\begin{eqnarray}
xF_T^D(x,Q^2) &=& \int\frac{\ud^3\bm{p}}{(2\pi)^3}
\left|\Psi_D(\bm{p})\right|^2\left(1+\frac{\gamma p_z}{M}\right) \\ \nonumber
 &&\left(x'F_T^{N}(x',Q^2;p^2)
+\frac{{x'}^2\bm{p}^2_\perp}{Q^2}F_2^{N}(x',Q^2;p^2)\right),\\
\gamma^2 F_2^D(x,Q^2) &=& \int\frac{\ud^3\bm{p}}{(2\pi)^3}
\left|\Psi_D(\bm{p})\right|^2\left(1+\frac{\gamma p_z}{M}\right) \\ \nonumber
 &&\left(1+\frac{4{x'}^2(p^2+\frac32\bm{p}^2_\perp)}{Q^2}\right)
F_2^{N}(x',Q^2;p^2),
\end{eqnarray}
\end{subequations}
where $F_{2,T}^N = (F_{2,T}^p + F_{2,T}^n)/2$. 
The deuteron wave function $\Psi_D(\bm{p})$ squared describes the
probability to find the bound proton (or neutron) with momentum $\bm{p}$,
$x'=Q^2/2p{\cdot} q$ is the Bjorken variable of the bound nucleon with the
four-momentum $p$, which is given by the difference of the target
four-momentum and the four-momentum of the spectator nucleon.
Eqs.(\ref{D-SF}) are written for the target rest frame and the $z$ axis is
chosen such that the momentum transfer,
$q=(q_0,\bm{0}_\perp,-|\bm{q}|)$. In this reference frame
$p=(M_D-\sqrt{\bm{p}^2+M^2},\bm{p}_\perp,p_z)$, where $M_D$ and $M$ are the
deuteron and the nucleon mass respectively.
The kinematical factors in Eqs.(\ref{D-SF}) result from the projection
of the structure functions from the hadronic tensor.
The transverse motion of the bound nucleon in the deuteron rest frame
is the reason for the appearance
of additional terms in the transverse and the longitudinal cross sections,
i.e. the terms proportional to ${x'}^2\bm{p}_\perp^2 F_2$ in Eqs.(\ref{D-SF}).
%
\subsection{Off-shell effects}
The bound proton and neutron are off-mass-shell and
their structure functions differ from those of the free proton
and neutron. 
The off-shell nucleon structure functions depend on the nucleon
virtuality $p^2$ as an additional variable. 
Therefore, off-shell effects in
the structure functions are closely related to the target mass corrections.
Target mass effects in the off-shell nucleon can be of two different kinds. 
First, similarly to the
on-shell nucleon, we have to take into account the kinematical target mass
dependence due to the finite $p^2/Q^2$ ratio. We assume that this
effect is described by Eqs.(\ref{TMC}), where
the nucleon mass squared is replaced by $p^2$ (this leads in turn to the
modification of the parameter $\gamma$ and the variable $\xi$ in the off-shell
region).
Furthermore, the dependence on $p^2$ appears already at  
leading twist 
\cite{GL92,KPW94,Ku98,MelSchTho}.
In order to estimate the off-shell effect in the LT structure functions,
we consider the quark distribution 
in terms of the spectral representation \cite{GL92,KPW94,Ku98}
\begin{eqnarray}\label{q:spec:off}
q(x,p^2)&=&\int\ud s\int\limits^{\lefteqn{\scriptstyle\ksqmax(x,s,p^2)}}
\ud k^2 \rho(s,k^2,x;p^2),\\
\ksqmax(x,s)&=&x\left(p^2-\frac{s}{1-x}\right).
\end{eqnarray}
The integration in \eq{q:spec:off} is taken over the mass spectrum of spectator
states $s$ and quark virtuality $k^2$, $\ksqmax$ is the kinematical
maximum of $k^2$ for the given $s$ and $p^2$. The invariant spectral
density $\rho$ measures the probability to find in a nucleon with
momentum $p$, a quark with light-cone momentum $x$ and virtuality $k^2$
and the remnant system in a state with invariant mass $s$.

We observe from \eq{q:spec:off} that the $p^2$ dependence of quark
distributions has two primary sources: the one in the upper
limit of $k^2$ integration (kinematical off-shell dependence), and
an explicit $p^2$ dependence of the quark spectral function $\rho$
(dynamical off-shell dependence).

The kinematical off-shell effect causes a negative correction to the bound
nucleon structure functions and produces an enhanced EMC effect, as
first noticed in \cite{GL92,KPW94}. However, if only the kinematical
off-shell effects are taken into account the number of valence quarks
in the nucleon would change with $p^2$. It can be seen directly from
\eq{q:spec:off} that the normalization of the quark distribution
decreases as $p^2$ decreases, provided that the spectral density is
positively defined.  This effect leads to an overall $1-2 \%$
depletion of valence quarks in the deuteron.  Furthermore, the
magnitude of this effect increases in heavy nuclei, since the average
shift from the mass shell of the bound nucleon increases.  This
observation indicates that an off-shell effect of dynamical origin
must also be present.

Dynamical off-shell effects can be viewed as a measure of the nucleon's
deformation inside the nuclear medium.
One possible way to evaluate dynamical off-shell effects is to require the
conservation of the valence quark number in the nucleon also in the off-shell region
\cite{KPW94,Ku98}:
\begin{equation}
\frac{\ud}{\ud p^2} \int_0^1\ud x\,  q_{\mathrm{val}}(x;p^2)=0,
\label{sr:conservation}
\end{equation}
This equation makes it possible to estimate
the off-shell effect minimizing the model dependence \cite{KPW94}.
It was shown that \eq{sr:conservation} results in a partial
cancellation between kinematical and dynamical off-shell effects \cite{KPW94,Ku98}.
However, the off-shell effect in the structure functions 
remains an important correction.
\subsection{Results and comparison with other approaches}

\begin{figure}
\begin{center}
%
\includegraphics{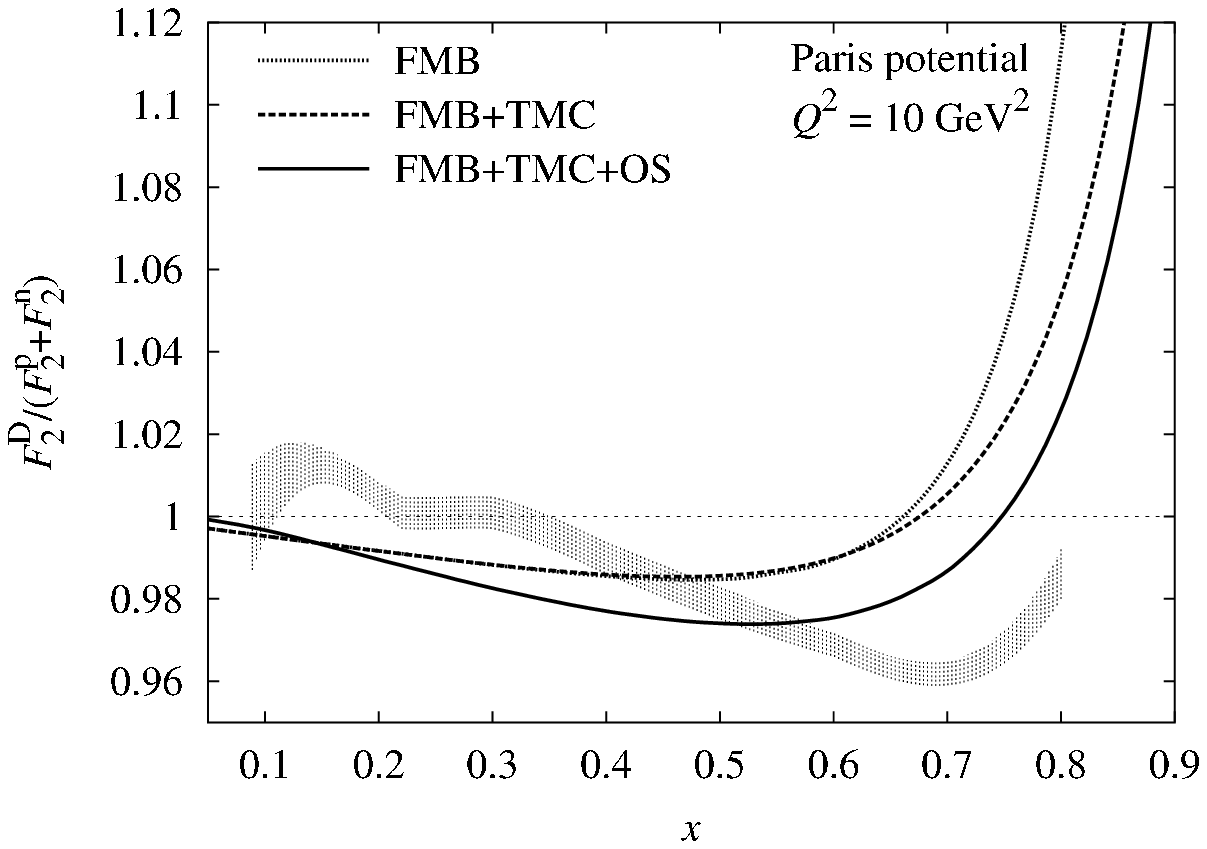}
\includegraphics{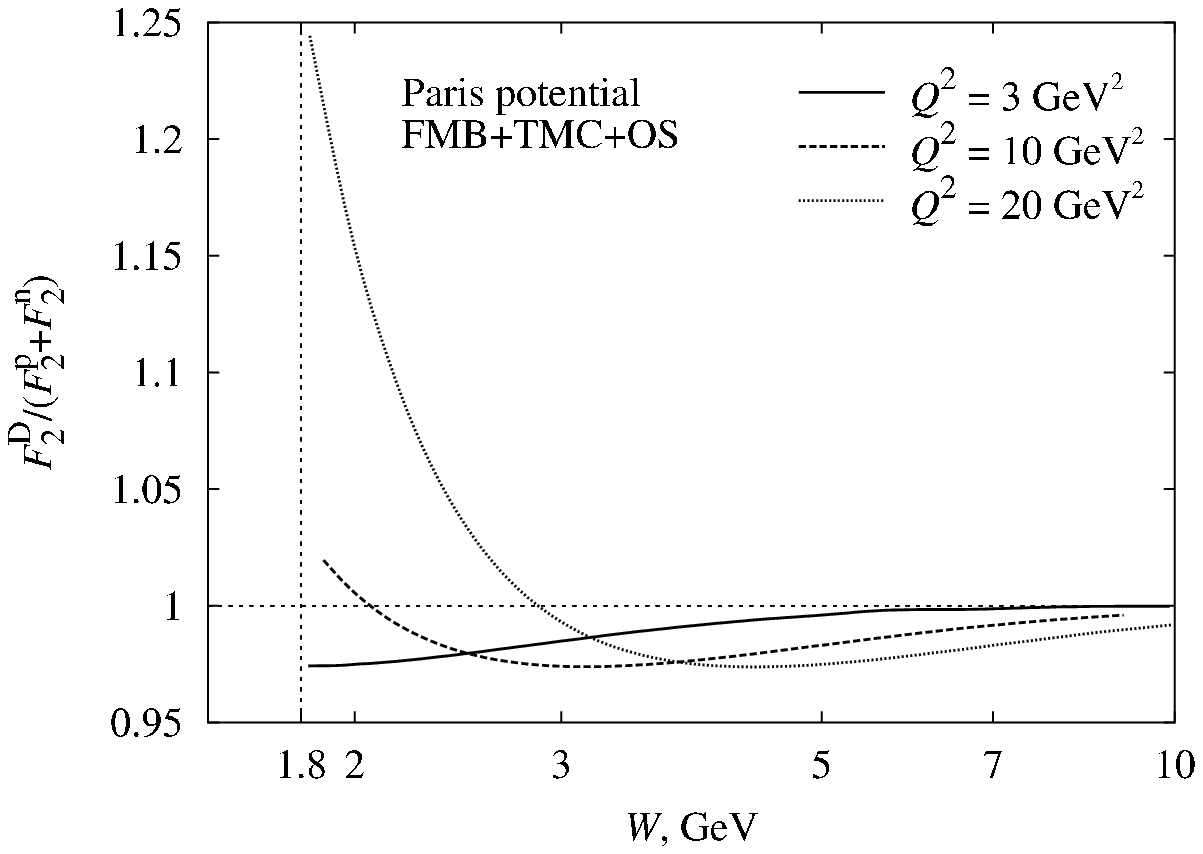}
\end{center}
\caption{The ratio $R_2^D$ calculated in different approximations.
In the upper panel this ratio is presented as a function of $x$ for fixed $Q^2$:
Fermi motion and binding effects (dotted line),
Fermi motion and binding effects combined with target mass corrections (dashed line); 
the full calculation including Fermi motion, binding, 
target mass and off-shell corrections is given by the solid line.
The shaded area in the upper panel corresponds to the prediction of the
nuclear density model of Ref.\cite{Gomez}.
In the lower panel the ratio $R_2^D$ is shown as a function of $W$ for a few different $Q^2$.
}
\label{r2d}
\end{figure}

The effect of nuclear corrections is illustrated in Fig.\ref{r2d}. The ratio
$R_2^D(x,Q^2)=F_2^D(x,Q^2)/F_2^N(x,Q^2)$ was calculated by Eqs.(\ref{D-SF})
using different approximations.
The dotted curve corresponds to the standard assumption
that the bound proton and neutron structure functions in Eqs.(\ref{D-SF})
are equal to those of the free nucleon ones. 
The competition between nuclear binding and Fermi motion determines
the shape of $R_2$. In particular, the depletion of nuclear structure functions
at $x<0.7$ is due to the effect of nuclear binding,  
while the rise of the $R_2$ ratio at large $x$
is due to the effect of the nucleon momentum distribution (Fermi motion) \cite{West74}.

The dashed curve corresponds to the results with the target mass
corrections.  We notice that TMC is an important correction at large
$x$, as can be directly seen from Eqs.(\ref{TMC}). This correction
modifies the shape and the magnitude of the LT structure function at
large $x$. This in turn leads to the softening of the ratio $R_2^D$ at
large $x$.

The solid curve stands for the full calculation with TMC and off-shell
effect taken into account. We observe, that the off-shell effect 
is most important in the binding region ($x$ between 0.3 and
0.7), where it causes a negative correction to the bound nucleon structure
functions.

The region of large $x$ corresponds to small masses of produced
hadronic states $W^2=M^2+Q^2(1/x-1)$.  For instance, the events with
$Q^2=10$\,GeV$^2$ and $x>0.75$ fall into the resonance region. For
this reason the DIS parametrization of structure functions, which are
used in computing the nuclear effects with \eq{D-SF}, are
questionable at large $x$. In order to avoid the resonance region and
elucidate nuclear effects in the DIS regime, we apply a cut at
$W=1.8$\,GeV.  The ratio $R_2^D$ as a function of $W$ was calculated
for a few different values of $Q^2$. The results are shown in the lower panel of
Fig.\ref{r2d}.

The off-shell effect is much less important in
the deuteron than in heavy nuclei. The strength of this effect is
governed by the average off-shellness of the bound nucleon
$\Delta=\langle p^2-M^2\rangle/M^2
\approx 2(\ceps- T)/M$, where
$T=\langle\bm{p}^2\rangle/2M$ and $\ceps=\langle p_0\rangle-M$ are the
average kinetic and separation energy. In order to illustrate the
strength of nuclear binding and off-shell effects, we have calculated
$T,\ \ceps$, and $\Delta$ averaged over the nuclear spectral function
for a number of nuclei.
In Fig.\ref{offness}, these parameters
are plotted as a function of the nuclear masss number $A$.
It can be seen that $\Delta$ increases by a factor of five
when going from the deuteron to heavy nuclei.

A phenomenological model of the EMC effect in the deuteron
was given in Ref. \cite{Gomez}. The model is based on an extrapolation of 
SLAC data on the EMC effect for a number of nuclei from $^4$He to
$^{197}$Au, where the key assumption was made that the quantity
$(F_2^A/F_2^N-1)$ -- $F_2^A$ being the nuclear structure function per nucleon --
scales with the nuclear density:
\begin{eqnarray}\label{ndm}
\frac{F_2^D/F_2^N-1}{F_2^A/F_2^N-1}&=&\frac{\rho_d}{\rho_A},
\end{eqnarray}
where $\rho_d$ and $\rho_A$ are the number densities in the deuteron
and in a heavy nucleus (Nuclear Density Model (NDM) \cite{FS88}). It was
also assumed that this ratio is independent of $Q^2$. Using \eq{ndm},
the ratio $F_2^D/F_2^{N}$ was obtained in Ref.\cite{Gomez} 
in terms of the experimentally
measured ratio $F_2^A/F_2^D$ by using:
\begin{eqnarray}\label{ndm2}
\frac{F_2^D}{F_2^N} &\approx& 1+
\frac{\rho_d}{\rho_A-\rho_d}\left(\frac{F_2^A}{F_2^D}-1 \right),
\end{eqnarray}
and by averaging the quantity appearing on the r.h.s. over the 
SLAC nuclear data.  
The values of $F_2^D/F_2^N$ were given for $x$ corresponding to the 
data bins. These results are shown in  Fig.\ref{r2d} by a shaded region.

We observe that the NDM attempts to extrapolate ``density
scaling'' to the region of light nuclei, where the notion of density
is ill defined \cite{mabt00}. 
The value $\rho_d=0.024$ fm$^{-3}$ was used in Ref.\cite{Gomez},
which was derived using the r.m.s. radius of the deuteron. 
However, it is not clear what volume is occupied by the deuteron and
for this reason $\rho_d$ has a large ``theoretical'' uncertainty. 
Since the
quantity $F_2^D/F_2^N-1$ is proportional to $\rho_d$,
this theoretical uncertainty will directly translate
into an uncertainty for the extrapolated ratio $F_2^D/F_2^N$. This
was not given in Ref.\cite{Gomez} and it is likely to be larger than
the errors shown in Fig.\ref{r2d}.

\section{QCD Analysis and Fit}
\label{fit}
\subsection{Fitting Procedure}
\begin{figure}
\includegraphics[width=14cm]{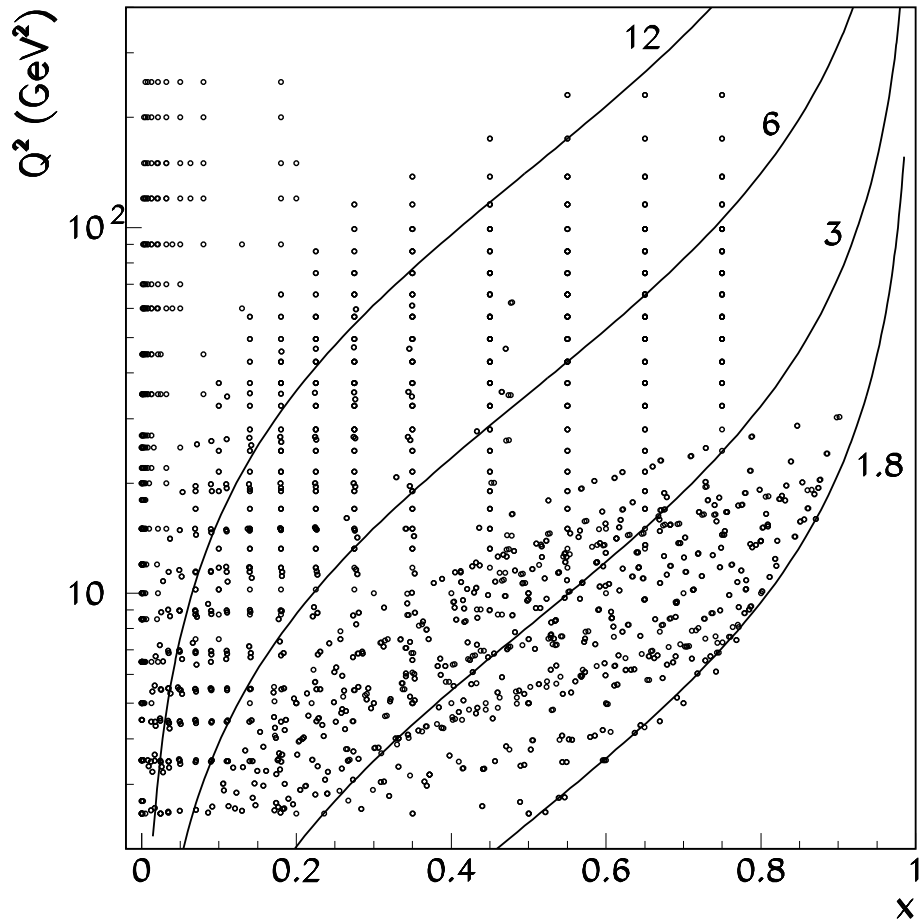}
\caption{Kinematic region of the data used in analysis. The
curves correspond
to constant values of the invariant mass $W$ whose values in units of GeV are
indicated in the plot.}
\label{fig:data}
\end{figure}

In our analysis the data on charged lepton DIS 
off proton and deuterium targets from  BCDMS, NMC, H1, 
ZEUS, and SLAC experiments were used~\cite{Whitlow:1992uw}. The cut 
$Q^2>2.5~{\rm GeV}^2$ was applied in order to avoid the region where 
$\alpha_s$ is large and the higher-order QCD radiative corrections 
can be out of control. The HERA data with $Q^2>250~{\rm GeV}^2$ were 
not used in the analysis since the impact of those data on the 
fit is marginal because of large experimental errors.
The maximum $x$ in the data set 
is about $0.9$ and minimum center-of-mass energy $W$ is about 
$1.8~{\rm GeV}$ (see Fig.\ref{fig:data}).
The total number of data points (NDP) is 1381 for the 
proton and 998 for the deuteron; $\chi^2/NDP=1.1$ for the best fit.

In our fit we make use of Eq.(\ref{xsect})
with the LT structure functions given in Eq.(\ref{LT_1})
corrected for the target mass effect by Eq.(\ref{TMC}).
The higher twist terms $H_{\rm 2,T}$ of Eq.(\ref{t-expansion}),
the parton distributions, and the value of $\alpha_{\rm s}$
were simultaneously fitted to the data.
The parton distributions were parametrized in the form used in the
earlier analyses of Ref.\cite{Ale1} 
with initial scale of the QCD evolution $Q_0^2=9~{\rm GeV}^2$.
The evolution equations were solved numerically by 
direct integration in $x$-space.
Our procedure is in agreement with the benchmarks introduced in  
Ref.~\cite{Giele:2002hx} that require the
precision of the solution to be much better than the 
accuracy of the data used in the analysis.

The implementation of target mass and nuclear corrections in our
fit is as follows.
The proton and neutron structure functions were calculated as 
\begin{eqnarray}\label{fit-tmc}
F_{\rm 2,T}^{p,n}=\mathcal{T}_{2,T}\left\{F_{\rm 2,T}^{p,n(\mathrm{LT})}
\right\}+{H_{\rm 2,T}^{p,n}}/{Q^2},
\end{eqnarray}
where we have expressed the target mass correction formula 
given in Eq.(\ref{TMC}) in terms of the functional: $\mathcal{T}_{2,T}$.

The deuteron structure functions were calculated as
\begin{eqnarray}\label{fit-D}
F_{\rm 2,T}^{D}&=&\mathcal{F}_{2,T}\left\{F_{\rm 2,T}^p+F_{\rm 2,T}^n\right\},
\end{eqnarray}
where $\mathcal{F}_{2,T}$ are the nuclear smearing functionals corresponding to
Eq.(\ref{D-SF}), and $F_{\rm 2,T}^{p,n}$ are the proton and neutron
structure functions of Eq.(\ref{fit-tmc}).
However, the
implementation of this approach slows down the numerical calculation because of 
4-dimensional integrations in the deuteron cross sections in terms of
the QCD-evolved PDFs.
We are able to reduce the calculation time
by using approximate expressions for the deuteron structure functions:

\begin{eqnarray}\label{fit-D2}
F_{\rm 2,T}^{D}&=&
\frac{ \mathcal{F}_{\rm 2,T}\left\{F_{\rm 2,T}^{p(QPM)}+F_{\rm 2,T}^{n(QPM)}
\right\}}
{F_{\rm 2,T}^{p(QPM)}+F_{\rm 2,T}^{n(QPM)}} \times \nonumber \\
& & \left[
\frac{\mathcal{T}_{\rm 2,T}\left\{F_{\rm 2,T}^{p(QPM)}+F_{\rm 2,T}^{n(QPM)}
\right\}}
{F_{\rm 2,T}^{p(QPM)}+F_{\rm 2,T}^{n(QPM)}} 
\left(F_{\rm 2,T}^{p(LT)}+F_{\rm 2,T}^{n(LT)}\right)
+\frac{H_{\rm 2,T}^{p,n}}{Q^2}
\right], 
\end{eqnarray}
where $F_{\rm 2,T}^{\rm QPM}$ is given by Eq.(\ref{fit-tmc}) with
the LT term in the ``quark-parton model'', that is without QCD 
radiative corrections to the
coefficient functions.
The deuteron structure
functions calculated using Eq.(\ref{fit-D2}) require 3-dimensional integration only.
This approximation introduces only a marginal bias in the final
results.  

We parametrize the functions $H_{\rm 2,T}(x)$ at the selected values
of $x$ and interpolate between those grid points using cubic splines.
The positions of the grid points were selected in such a way to
provide the overlap between the error-bars of nearest grid points.  The
values of the functions $H_{\rm 2,T}(x)$ at the grid points have been
fitted to data. 
This method made it possible to describe the different
structure functions in our fit and, at the same time, 
to keep the number of fitting parameters reasonable.

All experimental errors in the data including
uncorrelated statistical, correlated systematical, and
errors in overall normalizations have been taken into account
in the analysis using the
covariance matrix approach.  The error bands throughout the paper
are due to linear propagation of the errors into the fitted
parameters. In several of the experimental data sets used in our analysis the overall
normalization has been fixed by the authors from comparisons to
other data. 


For this reason we introduce an additional normalization
parameter for each target (in some cases also for each energy) in
the experiment and fit these parameters simultaneously with the
parameters of PDFs and the HT terms.
The overall normalization errors for such data subsets 
are accounted by the error propagation 
in the corresponding normalization factors through the general
Hessian matrix of the fit. Such treatment of the normalization errors 
allows for a maximal 
self-consistency of the analysis, incorporating all existing information 
and minimizing the normalizaton uncertainties in the data.
The re-normalized data subsets and
their re-normalization factors, derived in the fit, are listed in
Table~\ref{tab:norm}.
The re-normalization factors are generally close to 1
for the deuteron data and somewhat higher for the SLAC proton data. 
\subsection{Results}
\begin{figure}
\includegraphics[width=12cm]{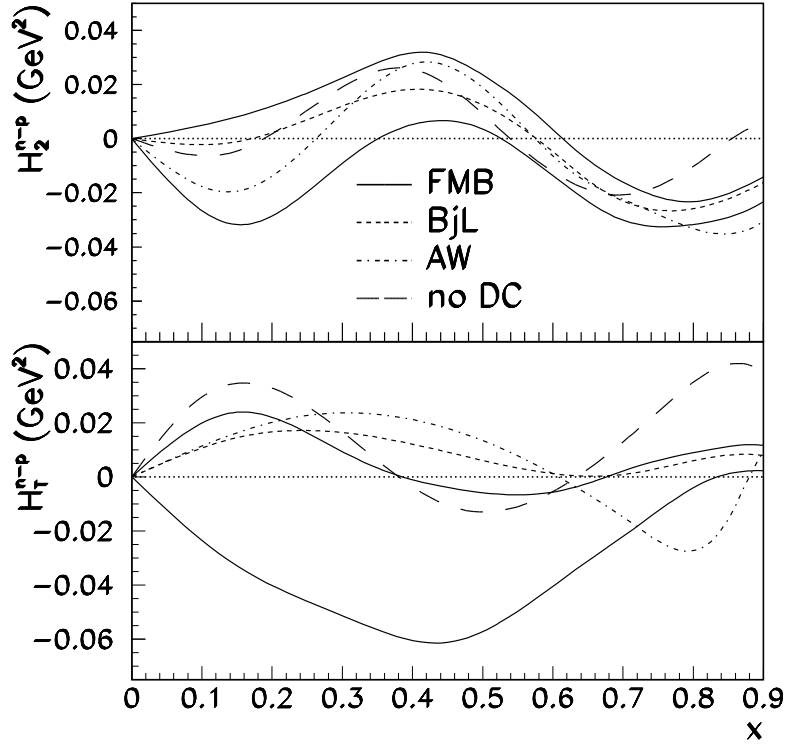}
\caption{Isospin asymmetries of
the HT terms obtained using different treatments 
of Fermi motion and binding corrections:
Eqs.(\ref{D-SF}) (delimited by solid lines, see text);
Atwood--West \protect\cite{West74} (dash--dotted line);
Eqs.(\ref{D-SF}) in the Bjorken limit, {\it i.e.} if all
$1/Q^2$ terms were disregarded (short dashes).
The curve with long dashes shows the result without 
Fermi motion and binding corrections.}
\label{fig:ia}
\end{figure}

\begin{figure}
\includegraphics[width=14cm]{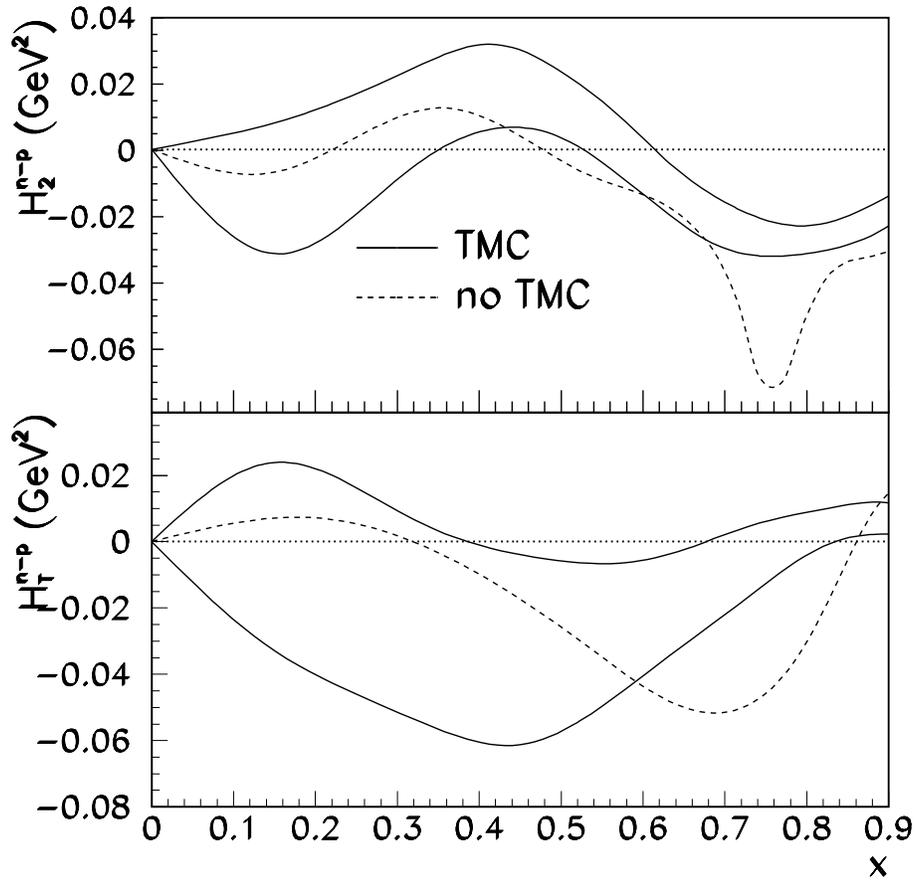}
\caption{Sensitivity of the isospin asymmetries to different 
approximations used in the calculation of 
the deuteron correction. Dashed line: fit with 
no TMC in the calculation of the deuteron correction;
Area between solid lines: fit with full treatment of the TMC).}
\label{fig:tmc}
\end{figure}

\begin{figure}
\includegraphics[width=12cm]{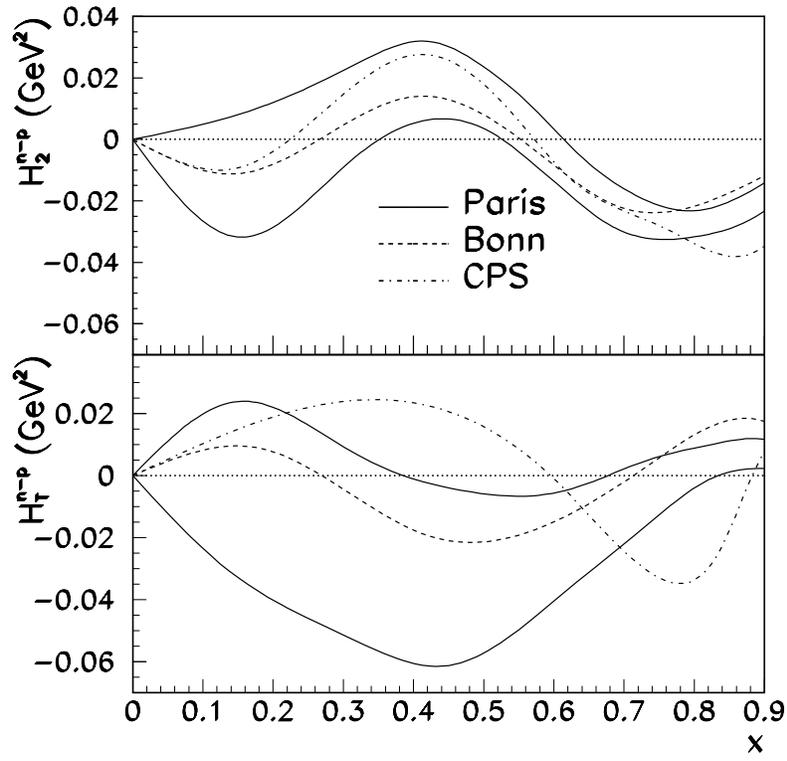}
\caption{Results of the fits with different deuteron 
wave functions: Ref.\protect\cite{Paris} (area between solid lines); Ref.\protect\cite{Bonn} 
(dashed line); Ref.\protect\cite{y-scaling} 
(dashed-dotted line).}
\label{fig:wf}
\end{figure}

\begin{figure}
\includegraphics[width=12cm]{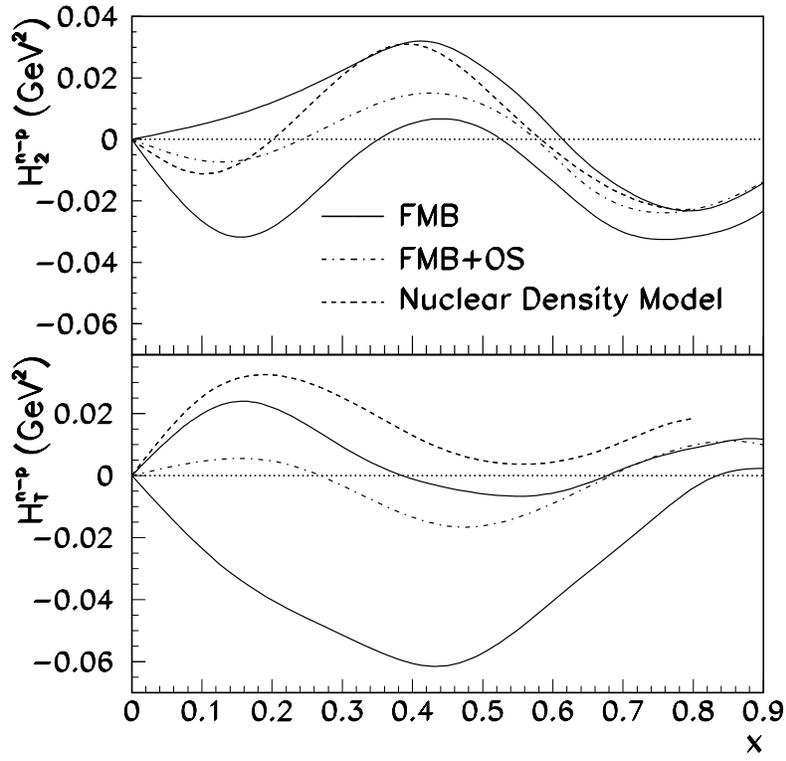}
\caption{Isospin asymmetries in the HT terms
obtained within the incoherent scattering approximation
(area between solid lines) as compared to the results obtained
with the treatment of off-shell
corrections (dashed-dotted lines). Results of the analysis based on the
nuclear density model of Ref.\protect\cite{Gomez}
are given by the dashed lines.}
\label{fig:off}
\end{figure}

\begin{figure}
\includegraphics[width=14cm]{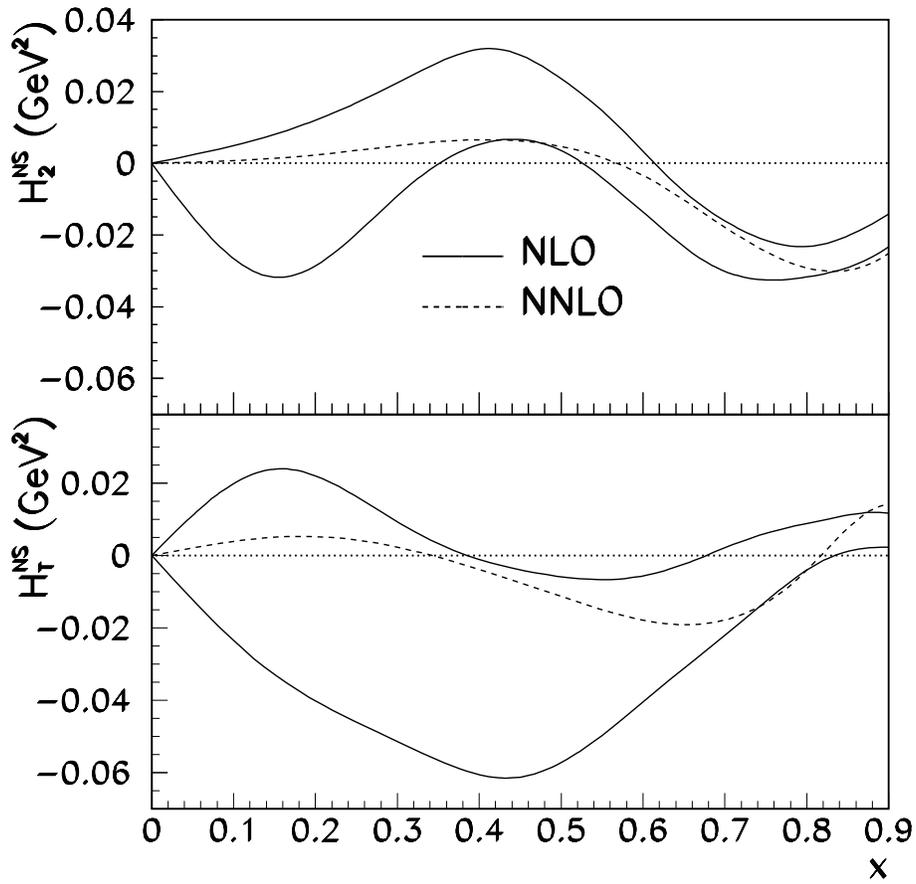}
\caption{Impact of NNLO QCD corrections on the isospin asymmetries
of the HT terms: NLO QCD fit (area between solid lines); NNLO QCD fit (dashed lines).}
\label{fig:nnlo}
\end{figure}

The isospin asymmetries of the HT terms in
$F_{\rm 2,T}$ obtained from this fit are shown in 
Figs.~\ref{fig:ia} to \ref{fig:nnlo}
as full lines, which correspond to the $1\sigma$ bands of 
the total experimental errors \cite{Ale-tables}.
In order to study the sensitivity of the results to various
theoretical assumptions and estimate theoretical errors, we performed
a number of fits using different approximations and assumptions.  

The
central value results of these fits are also given 
in Figs.~\ref{fig:ia} to \ref{fig:nnlo}.
Since the errors in $H_{\rm 2,T}$ obtained in all variants of the fit are 
correlated, the statistically significant shifts are those whose central 
values lie out of the bands given by the solid curves. 

For all variants of the fit we put the constraint $H_{\rm 2,T}(x)=0$
as $x\to 0$,
since the analyzed data are not sensitive to the isospin asymmetry 
at small $x$. Relaxing this constraint does not improve the 
quality of the fit and in this case
the values of $H_{\rm 2,T}(0)$ are
comparable to zero within errors. 

The sensitivity of the isospin asymmetry of the HT terms to the treatment
of Fermi Motion and nuclear Binding (FMB) in the deuteron 
is shown in Fig.\ref{fig:ia}.
One observes that different approaches 
lead to variations in $H_{2}^{n-p}$ up to several standard deviations
at $x \gtrsim 0.75$. 
The high sensitivity of the fit to details of the treatment of
Fermi motion indicates
that an accurate account of nuclear smearing 
in deuterium is crucial for the determination of 
$H_{\mathrm 2, T}^{n-p}$ at large $x$.

TMC also strongly affect the extraction of the isospin 
asymmetry due to their interplay with nuclear corrections
at large $x$, as it can be seen from Fig.\ref{r2d}. This is because TMC modify
the $x$ dependence of the LT structure functions at large $x$.  
The impact of TMC on the deuteron correction is illustrated in 
Fig.\ref{fig:tmc}, in which the fit
without TMC in $F_{\rm 2,T}^{p,n(QPM)}$ in Eq.(\ref{fit-D2}) is
compared to the fit with the full treatment of TMC. We observe a
noticeable difference between these two fits in the region of $x \approx 0.75$, 
especially for $H_{2}^{n-p}$.

We have also studied the sensitivity of our result to the 
choice of the deuteron wave function in Eq.(\ref{D-SF}). 
Calculations of deuteron wave functions predict different amounts of high momentum components  
depending on the nucleon--nucleon
potentials that are used, and on the treatment of 
relativistic effects.
We considered two extreme situations by comparing in Fig.\ref{fig:wf} results using the 
Bonn \cite{Bonn} and Paris \cite{Paris} potentials that 
predict the smallest high momentum components among modern calculations, 
and a phenomenological wave function that reproduces y-scaling 
data and that has therefore
a larger amount of high momentum components. Relativistic 
calculations seem also to have a larger amount of high momentum components 
\cite{bulgari}.  
One can see that the functions $H_{\rm 2,T}^{n-p}$ vary
within one standard deviation for the Paris and Bonn wave functions.
The phenomenological distribution, however, is more than one standard deviation
away at $x \geq 0.75$. 

\begin{figure}
\begin{center}
%
\includegraphics{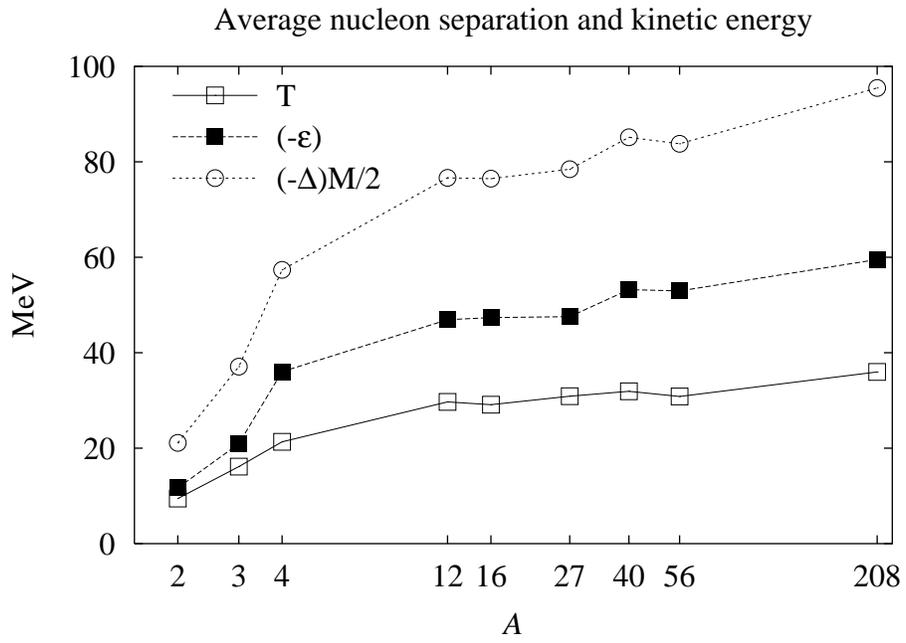}
\end{center}
\caption{
Average nucleon kinetic and separation energy and off-shellness $\Delta$
as functions of the
nuclear mass number $A$.}
\label{offness}
\end{figure}

The isospin asymmetry in the HT terms is also affected by 
off-shell effects in the bound nucleon structure functions
discussed in Section~\ref{nuclear-sect}.
Since the calculation of off-shell effects is model dependent, these
effects are in principle the main source of theoretical uncertainty 
in our analysis.
However, the shift from the mass shell for the 
bound proton and neutron
in the deuteron, measured by $\Delta=\langle p^2-M^2\rangle/M^2$, is small 
because of the weak binding in the deuteron
(see Fig.\ref{offness}). As a result, the net
effect of off-shell corrections is within the
$H_{\rm2,T}^{n-p}$ error-bars
(see Fig.\ref{fig:off}) and thus the uncertainty from the modeling
of off-shell effect is effectively small for the deuteron.
We note, however, that the off-shell effects will be much more important
in heavy nuclei, as also shown in Fig.\ref{offness}. 

In order to examine the impact of different models of nuclear effects 
in DIS, we repeated the fit with the deuteron correction calculated 
using the Nuclear Density Model (NDM) discussed in Section \ref{nuclear-sect}. 
Since the NDM correction is not available for $x>0.8$, we have removed the
corresponding data points from the fit. Furthermore, since the NDM
does not distinguish among structure functions, we assumed the correction 
to $F_{\rm T}$
to be the same as for $F_{2}$. The results are shown in Fig.\ref{fig:off}.  
The term $H_{\rm T}^{n-p}$ obtained from the NDM fit is different from the
corresponding term of the FMB fit, while for $H_{2}^{n-p}$ we observe a
good agreement between these two fits. However, since the nuclear density
model of the EMC effect in the deuteron is essentially different from the
approach discussed in this paper, this agreement appears to be
accidental.

Another source of theoretical uncertainty comes from higher order
QCD radiative corrections. These corrections decrease as $Q^2$
increases. For this reason the radiative corrections can simulate the
power-like terms in some kinematical regions of $Q^2$.  Indeed, it
is well known that the magnitude of phenomenological HT terms is
strongly correlated with the order of the QCD radiative corrections
applied in the analysis.  The HT terms drastically decrease when going
from LO to NLO in the structure function fit.  However, the magnitude
of the HT terms does not change much when going from NLO to NNLO, the
variation of the HT terms stays within one standard deviation
\cite{Ale1,Kataev:1999bp}. 
In the present analysis we also observe only a
marginal variation of $H_{\rm 2,T}^{n-p}$ after the NNLO corrections
have been applied (see Fig.\ref{fig:nnlo}).
We do not consider the soft gluon re-summation
\cite{ScScSt} as well as the log-like dependence of $H_{\rm 2,T}$
because of anomalous dimensions \cite{JiU}. The isoscalar part of
the HT terms can be affected by these effects at large $x\sim0.9$.
However, the isovector combinations $H_{\rm 2,T}^{n-p}$ seem to be much
less sensitive to these effects.

\begin{figure}
\includegraphics[width=14cm]{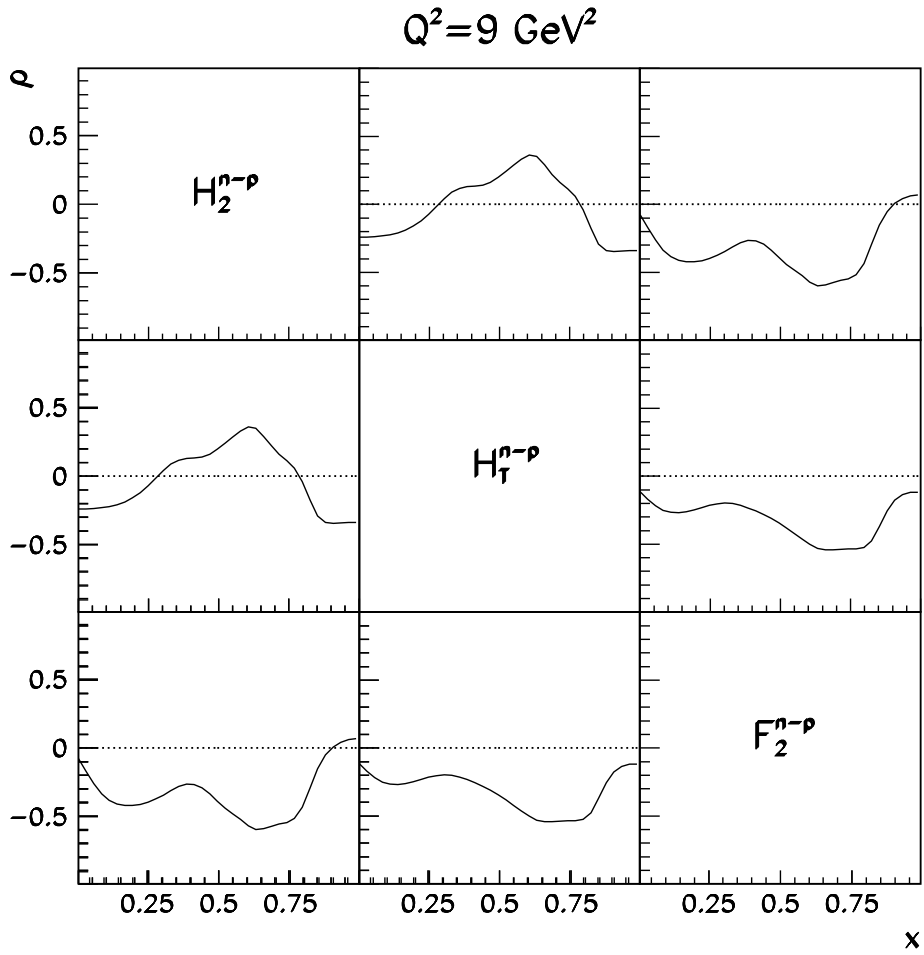}
\caption{The statistical 
correlation coefficients $\rho$ for the HT and LT terms
determined from our fit.}
\label{fig:cor}
\end{figure}

To summarize, the isospin asymmetry is stable with respect to the
theoretical uncertainties addressed in this paper. 
The statistical correlations between the LT and the HT terms in our
fit are illustrated in Fig.\ref{fig:cor}. A wide kinematical region of
considered DIS data allows to reliably separate the LT and HT
terms and, as a result, 
the corresponding correlation coefficients are less than 0.5.

\section{Phenomenology}
\label{pheno-sect}
\begin{figure}
\includegraphics[width=14cm]{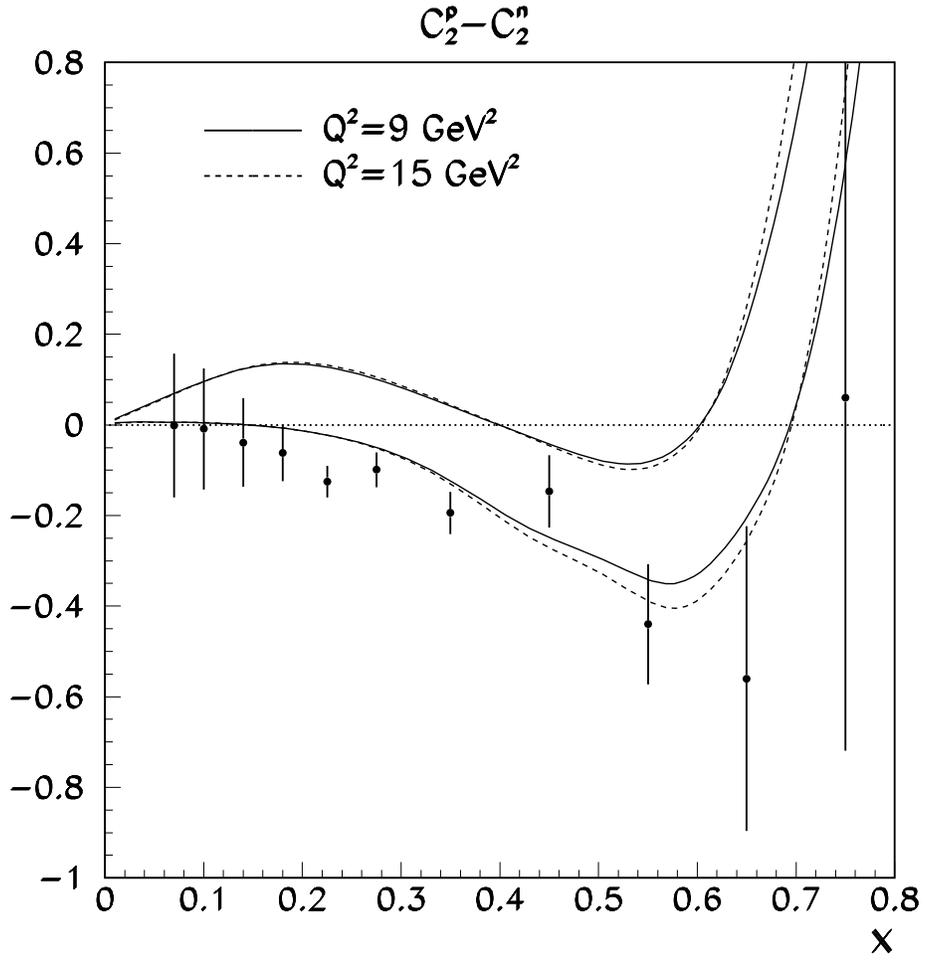}
\caption{The isospin asymmetry of the HT coefficients 
determined by NMC (points with errors bars) compared to
to our determination (area inside error bands).}
\label{fig:nmc}
\end{figure}

It is instructive to compare our results to the earlier extraction of
the isotopic effects in the HT terms.
The NMC extraction of the HT asymmetry in $F_2$ \cite{NMC92}
is based on the equation
$$
\frac{F_2^D}{F_2^p}-1=\frac{\mathcal{T}_2\{F_2^{n(LT)}\}}
{\mathcal{T}_2\{F_2^{p(LT)}\}}
\left(1-\frac{C_2^p-C_2^n}{Q^2}\right),
$$
in which the combined SLAC-NMC-BCDMS data for the deuteron and proton
are used.
The relation between the functions $C_2^{p,n}$ and the functions
$H_2^{p,n}$ from our analysis can be written as follows
\begin{equation}
C_2^p-C_2^n \approx \frac{H_2^p}{\mathcal{T}_2\{F_2^{p(LT)}\}}
-\frac{H_2^n}{\mathcal{T}_2\{F_2^{n(LT)}\}}.
\label{eqn:nmc}
\end{equation}
It must be noted, that 
the correspondence between these two 
definitions is somewhat uncertain since the denominators in
Eq.(\ref{eqn:nmc}) depend on $Q^2$.  
However for comparison, in Fig.\ref{fig:nmc} we plot the difference $C_2^p-C_2^n$
calculated from our results using Eq.(\ref{eqn:nmc}) at different
$Q^2$. Both extractions agree within errors. 
We note that our errors are systematically smaller because of a larger number of
data points used in the analysis.
It must be also emphasized that our analysis and the one
by NMC are different in a few aspects (no deuteron corrections were
applied in the NMC analysis, different treatments of systematic
errors were used, etc.). All of these factors could be responsible for 
the remaining discrepancies.

\begin{figure}
\includegraphics[width=14cm]{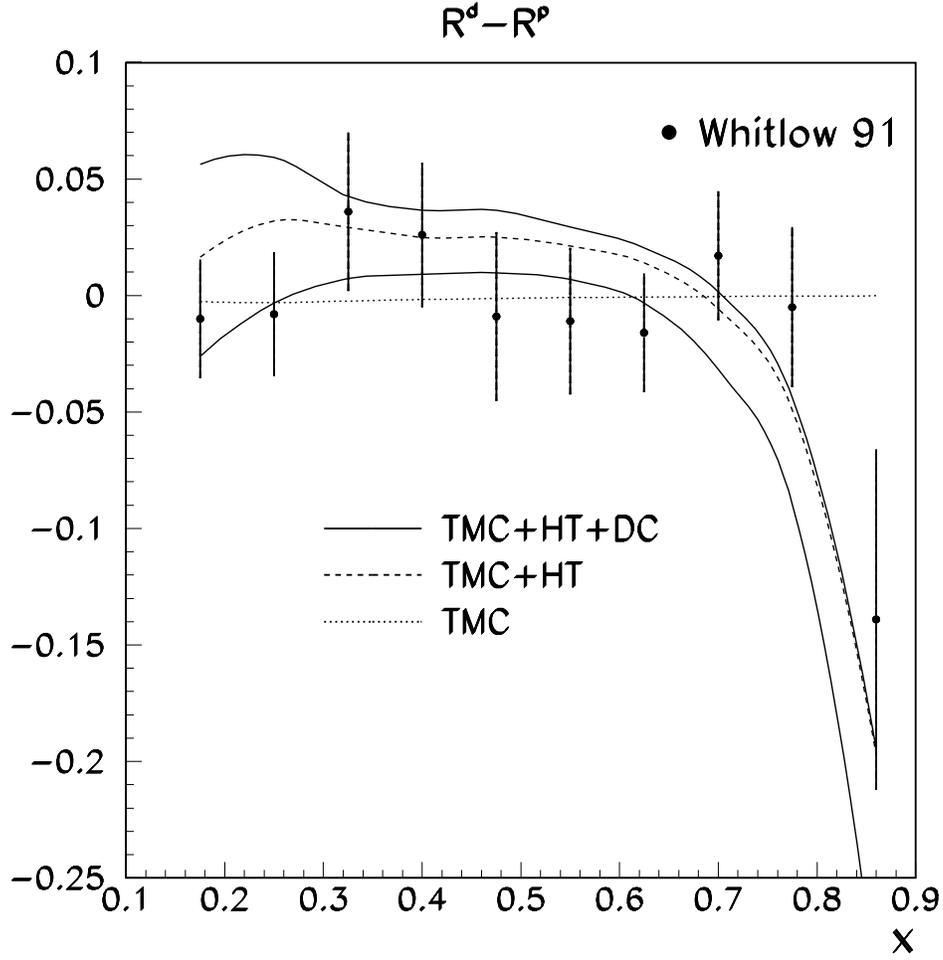}
\caption{The isospin asymmetry in the structure function 
$R$ determined from the SLAC data (points with error bars) compared to
to our results. The LT contribution corrected for the TMC 
with no deuteron correction (dotted line);
both the LT and HT terms but no deuteron corrections (dashed line);
the same as above with the deuteron correction (area between solid lines).}
\label{fig:rslac}
\end{figure}

In Fig.\ref{fig:rslac}, the isospin asymmetry in the
ratio $R=\sigma_L/\sigma_T$ 
$$
R=\gamma^2\frac{F_2}{F_T}-1,
$$
extracted from the fit of the SLAC data~\cite{Whitlow:1990gk},
is compared to our results. 
A good agreement on $R^D-R^p$ suggests that our fit is self-consistent,
since our analysis includes the data used in Ref.\cite{Whitlow:1990gk}.
The value of $R^D-R^p$ at $x=0.03\div0.35$ measured by NMC
\cite{Arneodo:1996kd} is also comparable to the calculations based on our fit.
In Fig.\ref{fig:rslac} we show the different contributions to 
$R^D-R^p$ considered in our analysis. It can be seen that 
at large $x$ the value of $R^D-R^p$ is defined mainly by the HT terms.

\begin{figure}
\includegraphics[width=14cm]{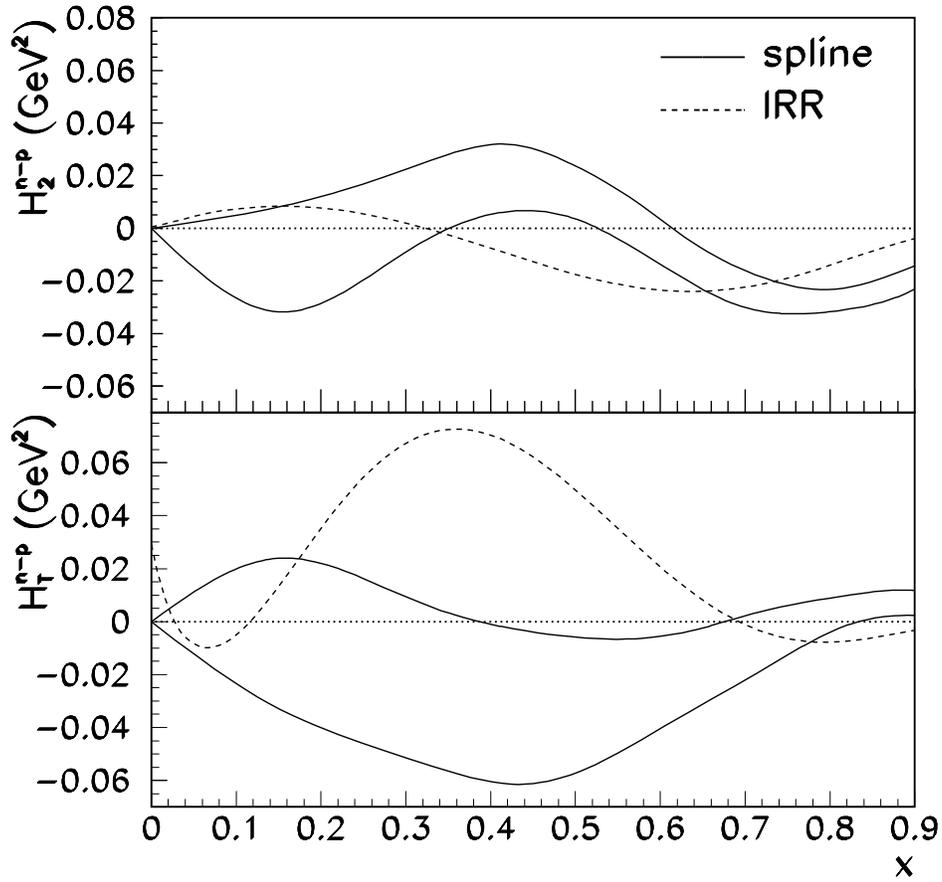}
\caption{Comparison of the phenomenological  
isospin asymmetries (area between solid lines) to the prediction of the 
infrared renormalon model (dashed lines).} 
\label{fig:model}
\end{figure}

\begin{figure}
\includegraphics[width=14cm]{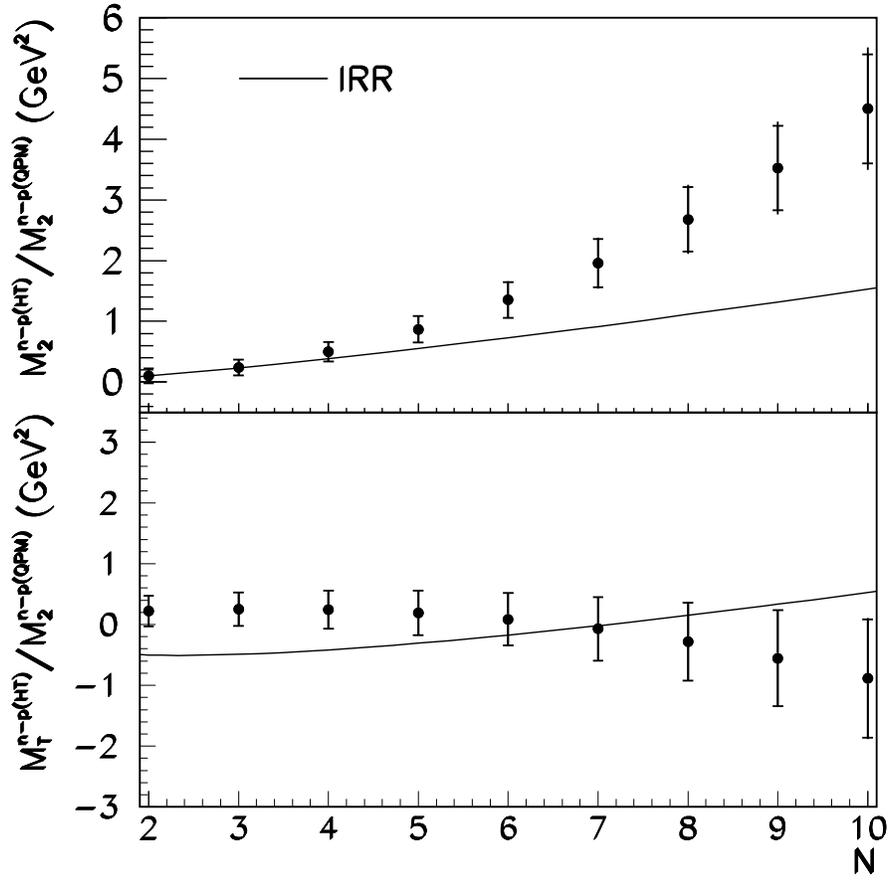}
\caption{The ratios of the Mellin moments of HT terms to the ones of 
LT terms (points with error bars) 
are compared with predictions of the IRR model (area between solid lines).
The inner bars give total experimental errors in the moments, 
the outer bars include the error due to the extrapolation into 
the unmeasured region at $x>0.9$.}
\label{fig:htmom}
\end{figure}

The isotopic asymmetries $H_{\rm 2,T}^{n-p}$ determined from our fit can be
also compared with the predictions of different theoretical models.  A
popular model is the infrared renormalon model (IRR) of
Refs.\cite{DMW,SMMS}. In this model HT terms derive from the
resummation of multi-loop diagrams, and their $x$ dependence 
is obtained as the Mellin convolution of the LT terms with 
flavor-independent coefficient functions $C^{\rm IRR}$, 
$$ 
H_{\rm 2,T}=A_2^\prime C_{\rm 2,T}^{IRR} \otimes F_2^{QPM}.
$$
The dimensional normalization factor $A_2^\prime$ determines the
characteristic scale of the HT terms.  This scale is not determined in
the IRR model and it is adjusted from comparisons with the data.  In
Fig.\ref{fig:model}, our results are compared to the IRR model
calculation for the non-singlet contribution, with the normalization
factor $A_2^\prime=-0.3~{\rm GeV}^2$.  It is clear that the
agreement is poor and it cannot be improved by simply varying the normalization
factor of the IRR model.
Alternatively, the comparison can be made in terms of the Mellin 
moments 
$$
M_{\rm 2,T}(N)=\int_0^1 dx x^{N-2} H_{\rm 2,T}(x).
$$
The Mellin moments of $H_{\rm 2,T}^{n-p}$ are given in  
Table \ref{tab:mom}.
In Fig.\ref{fig:htmom}, the ratio of these moments to the
moments of $F_2^{QPM}$ is plotted. Also shown are the
moments of $C_{\rm 2,T}^{IRR}$.  We only show $N\geq 2$,
since the data do not allow us to constrain the behavior of $H_{\rm
2,T}^{n-p}$ at small $x$.  Again, we observe a disagreement with the
IRR model.
Parameterizations of DIS structure functions interpolating
between the low and high $Q^2$ regions are also available \cite{CKMT,SzcUle}. 
The comparison of our results with these models is not directly feasible, 
because of the strong $Q^2$ dependence introduced in their evaluation of the 
HT terms, due to effects beyond the OPE.

\begin{figure}
\includegraphics[width=14cm]{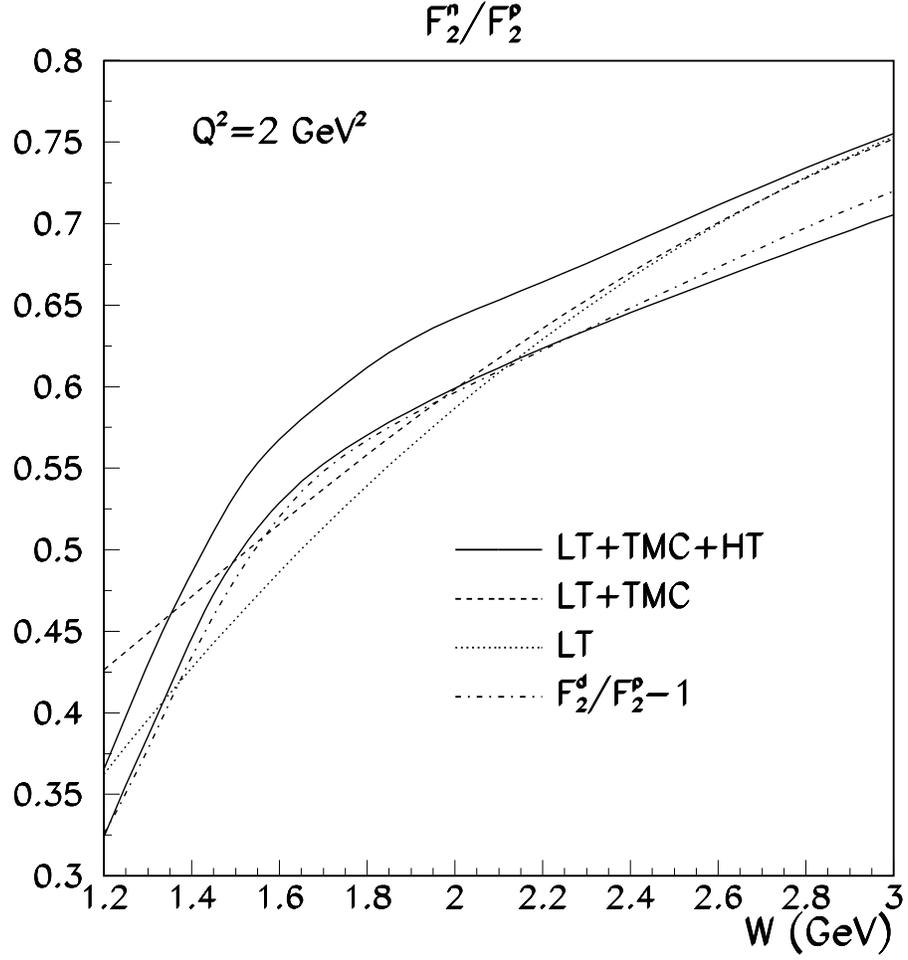}
\caption{The ratio $F_2^n/F_2^p$
calculated in different approximations and extrapolated 
to the resonance region: LT terms only
(dotted line); effect of TMC (dashed line);
TMC and HT terms (area between solid lines).  
}
\label{fig:f2np}
\end{figure}

\begin{figure}
\includegraphics[width=14cm]{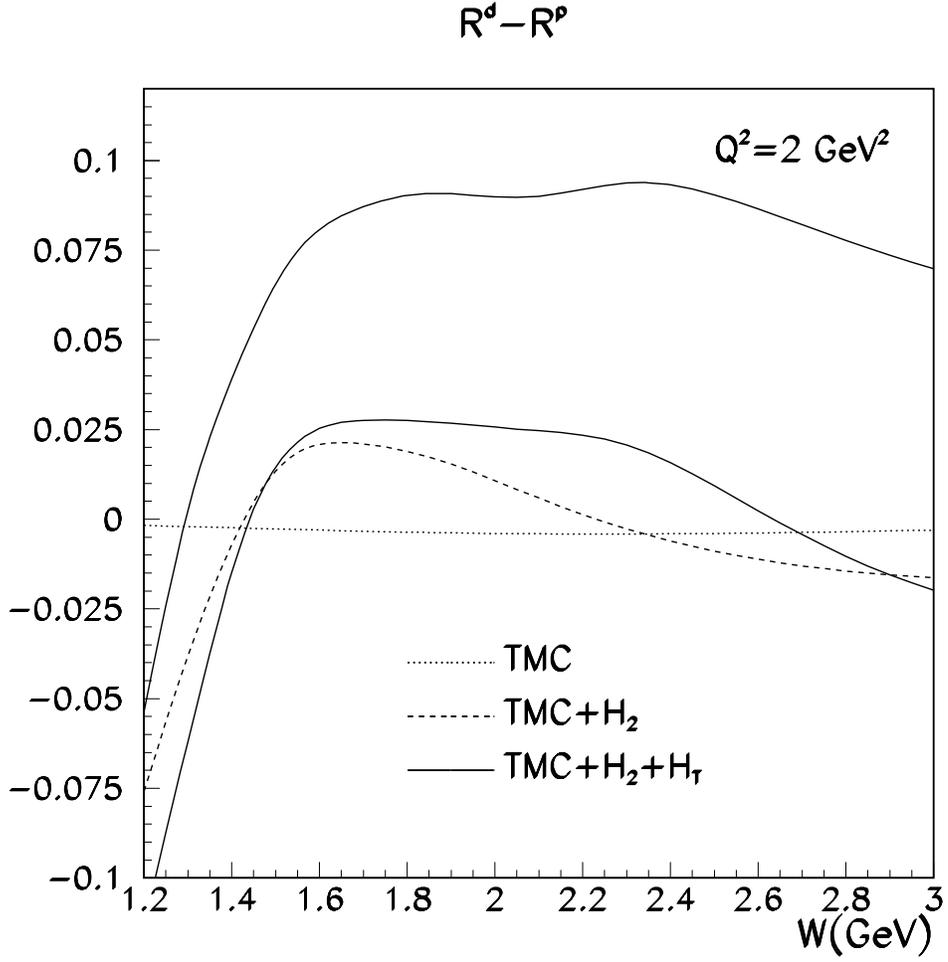}
\caption{The difference $R^D-R^p$ calculated in different approximations and
extrapolated to the 
resonance region: only the LT terms with the TMC (dotted line);
the same as above and the contribution from $H_{\rm 2}$ (dashed line);
the same as the dashed line and the contribution from 
$H_{\rm T}$ (area between solid lines).}
\label{fig:rjlab}
\end{figure}

Having estimated the power corrections to different structure functions,
one can make an extrapolation 
into the resonance region in order to study the 
phenomenon of duality ~\cite{Bloom:1970xb}. Examples of such extrapolations are given in 
Figs.\ref{fig:f2np},\ref{fig:rjlab}. At values
of $Q^2$ relevant for the experiments at Jefferson Lab the HT terms  
contribute moderately to the ratio $F_2^n/F_2^p$. For the difference 
$R^D-R^p$ the impact of the HT terms is larger. At $W\lesssim 1.4~{\rm GeV}$
the dominant contribution comes from $H_2^{n-p}$. At  
$W\gtrsim 1.4~{\rm GeV}$ the contribution from  $H_2^{n-p}$ is small
and the main effect comes from $H_{\rm T}^{n-p}$. However the errors are 
large in this region, since it corresponds to $x\sim 0.4$, where 
$H_{\rm T}^{n-p}$ has its largest uncertainty. 

%
%

\section{Conclusions}
\label{conclusions}
In conclusion, we determined the
isospin asymmetries of the functions $H_{\rm 2}$ and $H_{\rm T}$ 
describing the HT terms of the 
DIS structure functions. The value of $H_{\rm T}$ is
consistent with zero within the errors for all values of $x$.
Also $H_{\rm 2}$ is consistent with zero at low and intermediate values of $x$;
It deviates from zero at $x \gtrsim 0.7$. 
We performed a careful study of the theoretical uncertainties that might affect the 
extraction and we conclude that they 
do not overwhelm the effect. 
The asymmetry $H_{\rm 2}^{n-p}$ 
is negative at large $x$. It reaches its maximum at $x \approx 0.8$ 
where it is $\sim 0.03~{\rm GeV}^2$, 
in agreement with the order of magnitude 
of the scale of QCD, $\Lambda^2$. 
We also find that the
$x$ dependence of $H_{\rm 2,T}^{n-p}$ is in poor agreement with the predictions
of the IRR model. 
For more conclusive comparisons 
more precise data at $x\sim 0.4$ and $Q^2\sim 1~{\rm GeV}^2$
are necessary.

\begin{acknowledgments}
We are indebted to S. Brodsky, E.~Gardi, A.~Vainshtein, 
for discussions. We also thank A.~Kaidalov and A.~Szczurek for comunications.
Two of us (S.A. and S.K.)
are grateful to the University of Virginia, where the final part of 
this work was done, for hospitality.
This work was supported by grants from the National Research Council COBASE program, from the U.S. Department
of Energy under grant no. DE-FG02-01ER41200, and from the Russian Foundation for Basic Research under grant no. 03-02-17177. 
\end{acknowledgments}


\newpage
\begin{table}[h]
\caption{\label{tab:norm}
List of re-normalized experiments with the corresponding re-normalization 
factors $\eta$.}
\begin{ruledtabular}
\begin{tabular}{ccc} 
Experiment&\multicolumn{2}{c}{$\eta$[\%]} \\ \cline{2-3} 
&proton &deuterium  \\  
SLAC-E-49A   &$1.5\pm1.3$  &$-1.1\pm1.2$   \\
SLAC-E-49B  &$2.8\pm1.3$&$0.2\pm1.2$  \\
SLAC-E-87   &$2.9\pm1.2$&$0.9\pm1.2$   \\
SLAC-E-89B  &$1.4\pm1.2$&$-1.1\pm1.2$   \\
SLAC-E-139   &&$0.6\pm1.3$  \\
NMC(90 GeV)  &$-0.8\pm1.4$&$-2.3\pm1.3$  \\
NMC(120 GeV) &$0.8\pm1.3$&$-1.4\pm1.3$ \\
NMC(200 GeV) &$2.4\pm1.3$&$0.1\pm1.3$ \\
NMC(280 GeV) &$1.2\pm1.3$&$-1.0\pm1.3$ \\
\end{tabular}
\end{ruledtabular}
\end{table}

\begin{table}[ht]
\label{tab:mom}
\caption{The Mellin moments of $H_{\rm 2,T}^{n-p}$.}
\begin{ruledtabular}
\begin{tabular}{ccc} 
N& $M^{n-p}_{\rm 2}(N)$ & $M^{n-p}_{\rm T}(N)$ \\ 
2 & $-0.0058\pm0.0069$ & $-0.012\pm0.014$ \\
3 & $-0.0046\pm0.0024$ & $-0.0048\pm0.0052$ \\
4 & $-0.0041\pm0.0013$ & $-0.0020\pm0.0026$ \\
5 & $-0.00355\pm0.00084$ & $-0.0008\pm0.0015$ \\
6 & $-0.00301\pm0.00061$ & $-0.00019\pm0.00096$ \\
7 & $-0.00254\pm0.00048$ & $ 0.00009\pm0.00068$ \\
8 & $-0.00215\pm0.00040$ & $ 0.00023\pm0.00052$ \\
9 & $-0.00183\pm0.00034$ & $ 0.00029\pm0.00041$ \\
10 & $-0.00157\pm0.00030$ & $ 0.00030\pm0.00034$ \\

\end{tabular}
\end{ruledtabular}
\end{table}

\newpage
%
%
%
%
%
%
%


\end{document}